\documentclass[final,1p,times,twocolumn]{elsarticle}

\usepackage{graphicx}
\usepackage{amssymb}

\journal{Nucl.\ Instrum.\ Methods Phys.\ Res.\ A}

\begin{document}

\begin{frontmatter}


\title{Initial Operation of the Recoil Mass Spectrometer EMMA at the ISAC-II Facility of TRIUMF}



\author[label1,label2]{B.~Davids}
\author[label1,label3]{M.~Williams}
\author[label1]{N.E.~Esker}
\author[label1]{M.~Alcorta}
\author[label1]{D.~Connolly}
\author[label3]{B.R.~Fulton}
\author[label1,label2]{K.~Hudson}
\author[label1]{N.~Khan}
\author[label1]{O.S.~Kirsebom}
\author[label1]{J.~Lighthall}
\author[label1]{P.~Machule}

\address[label1]{TRIUMF, 4004 Wesbrook Mall, Vancouver, BC, V6T 2A3, Canada}
\address[label2]{Department of Physics, Simon Fraser University, 8888 University Drive, Burnaby, BC, V5A 1S6, Canada}
\address[label3]{Department of Physics, University of York, Heslington, York, YO10 5DD, United~Kingdom}

\begin{abstract}
The Electromagnetic Mass Analyser (EMMA) is a new vacuum-mode recoil mass spectrometer currently undergoing the final stages of commissioning at the ISAC-II facility of TRIUMF. EMMA employs a symmetric configuration of electrostatic and magnetic deflectors to separate the products of nuclear reactions from the beam, focus them in both energy and angle, and disperse them in a focal plane according to their mass/charge $(m/q)$ ratios. The spectrometer was designed to accommodate the $\gamma$-ray detector array TIGRESS around the target position in order to provide spectroscopic information from electromagnetic transitions. EMMA is intended to be used in the measurement of fusion evaporation, radiative capture, and transfer reactions for the study of nuclear structure and astrophysics. Its complement of focal plane detectors facilitates the identification of recoiling nuclei and subsequent recoil decay spectroscopy. Here we describe the facility and report on commissioning efforts. 
\end{abstract}


\end{frontmatter}


\section{Introduction}\label{intro}
\begin{figure}[h]
\centering\includegraphics[width=\linewidth]{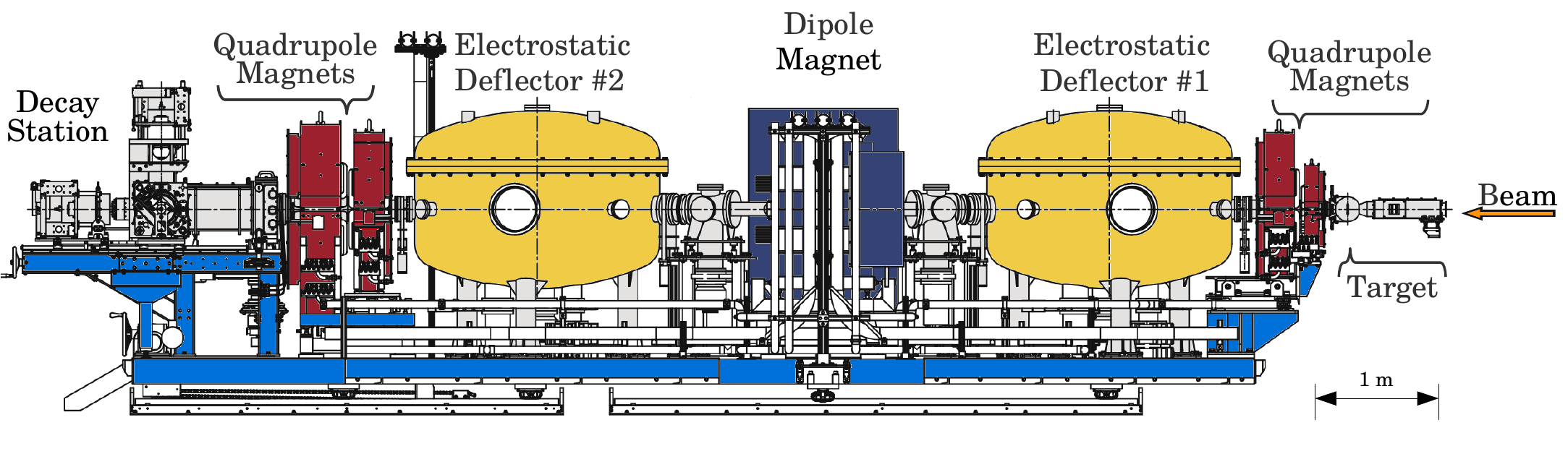}
\caption{\label{layout}Schematic side view of EMMA, showing the target chamber, quadrupole and dipole magnets,  electrostatic deflectors, and focal plane detector chamber surrounded by $\gamma$-ray detectors. The spectrometer is mounted on a platform capable of 1.5 m of travel along the beam direction.}
\end{figure}

\begin{figure}[h]
\centering\includegraphics[width=0.75\linewidth, angle=270]{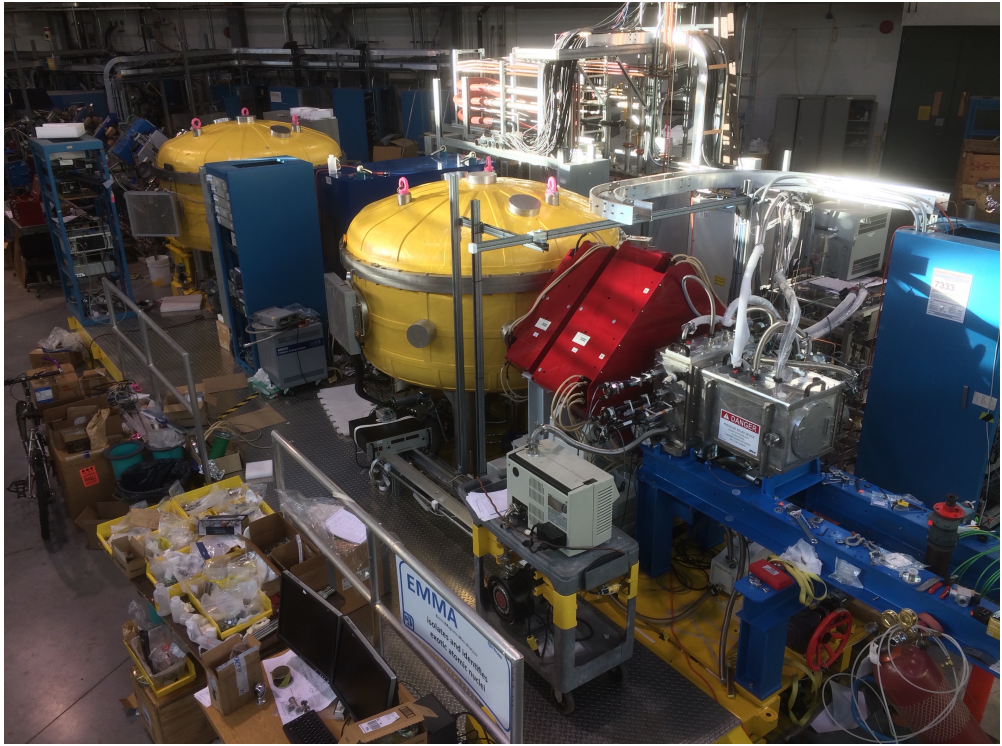}
\caption{\label{photo}Photograph of EMMA taken in December 2016.}
\end{figure}

The Electromagnetic Mass Analyser (EMMA) \cite{davids05} has been installed at the ISAC-II facility of TRIUMF \cite{laxdal14}. EMMA is a vacuum-mode recoil mass spectrometer designed to separate the recoils of nuclear reactions from the primary beam, focus them in energy and angle, and disperse them in a focal plane according to their mass/charge $(m/q)$ ratios. The spectrometer is fixed at 0$^\circ$ with respect to the beam axis and is mounted on a common support platform with 1.5 m of longitudinal travel, allowing for the positioning of various detector arrays at the target position, including the $\gamma$-ray spectrometer TIGRESS \cite{hackman14} and the Si charged particle detector array SHARC \cite{diget11}. As depicted in Figure \ref{layout}, EMMA uses a symmetric configuration of two electrostatic deflectors and a dipole magnet to focus reaction products in kinetic energy/charge $(E/q)$. Angular focusing is achieved via quadrupole doublets at the entrance and exit of the spectrometer, the latter of which enables variable $m/q$ dispersion. A photograph of the spectrometer taken in December 2016 is shown  in Figure \ref{photo}.

\section{Ion Optics}\label{optics}

The ion optical design of EMMA is similar to those of the Rochester RMS \cite{cormier81, cormier83}, CAMEL at Legnaro \cite{spolaore85, signorini94, spolaore95}, the Oak Ridge RMS \cite{cormier90, cole92, gross00}, HIRA at IUAC \cite{sinha94,nath07}, the JAERI RMS \cite{ikezoe96,ikezoe97,kuzumaki99}, and the Argonne Fragment Mass Analyzer (FMA) \cite{davids89,davids92,back96}; it was optimized to provide large acceptance without unduly compromising the resolving power necessary to study transfer reactions in inverse kinematics as well as fusion evaporation reactions. To that end, EMMA features electrodes with larger bending radii and a shorter first quadrupole magnet than the FMA. The EMMA electrodes have the same bending radii as those of HIRA and are smaller than those of the Oak Ridge RMS. For ions of a given electrostatic rigidity, larger electrode bending radii allow operation at lower voltages. The ion optics code GIOS \cite{wollnik87} was used to design the spectrometer, both initially and again after the electromagnetic elements were fabricated, to take account of the differences between their specified and as-built properties. The standard achromatic ion optical tune has a vertical crossover in the centre of the dipole magnet and $m/q$ dispersion $(x|\delta_m) \equiv \frac{\partial x}{\partial \delta_m}$ of 10 mm/\%. Here, $x$ is the horizontal displacement with respect to the optic axis in the focal plane and $\delta_m \equiv \frac{m/q-m_0/q_0}{m_0/q_0}$ is the fractional $m/q$ deviation with respect to that of the central reference trajectory, $m_0/q_0$. Two quadrupole doublets permit   variable angular focussing modes; by changing the fields in the second doublet, the $m/q$ dispersion can be varied continuously between 0 and 20 mm/\%.

To first order, given the bending angles and radii of curvature of the electrostatic deflectors and the dipole magnet as well as the edge angles of the latter, EMMA achieves energy focussing on account of the 1225~mm separation between the effective field boundaries of the dipole magnet and the electrostatic deflectors on either side of it. Mathematically this is expressed as $(x|\delta_E) \equiv \frac{\partial x}{\partial \delta_E} = 0$ and $(a|\delta_E) = 0$. The fractional kinetic energy/charge deviation with respect to that of the central reference trajectory, $E_0/q_0$, is defined as $\delta_E=\frac{E/q-E_0/q_0}{E_0/q_0}$ and $a \equiv p_x/p_0\approx\theta$, the horizontal angle with respect to the optic axis. Here $p_x$ is the horizontal projection of the ion momentum and $p_0$ is the total momentum of the reference trajectory. In the horizontal direction the angular acceptance is defined by the gap of the first electrostatic deflector and in the vertical direction it is defined by the vacuum chamber of the first quadrupole magnet. The standard distance from the target to the effective field boundary of the first quadrupole is 25~cm and the focal plane position is variable but in the standard tune lies 32 cm from the effective field boundary of the fourth quadrupole. 

\begin{table}
 \caption{\label{dimensions}As-Built EMMA Dimensions and Maximum Fields}
 \begin{tabular}{llll}
 \\
Length from target to focal plane (m) & 9.143 & &\\
 \hline
Dipoles & MD & ED1, ED2 &\\
\hline
Radius of curvature (m) & 1.0 & 5.0 &\\
Deflection angle ($^\circ$) & 40.11 & 20.05 &\\
Entrance and exit inclination angles ($^\circ$) & 7.93, 8.67 & --- &\\
Effective field boundary radii (m) & 3.472 &--- &\\
Pole gap (cm) & 12 & 12.5 &\\
Maximum field & 1.0 T & 40 kV/cm &\\
Maximum rigidity & 1.0 T m & 20 MV &\\
\hline
Magnetic lenses & Q1 & Q2, Q3 & Q4\\
\hline
Bore diameter (cm) & 7 & 15 & 20\\
Effective length (cm) & 14.0 & 30.0 & 40.2\\
Maximum pole tip field (T) & 1.21 & 0.84 & 0.80\\
Maximum field gradient (T/m) & 35 & 12 & 8.1\\
\hline
\end{tabular}
\end{table}


\section{Electromagnetic Elements}\label{em_elements}

The quadrupole  and dipole magnets, their power supplies, and most of the components of the electrostatic deflectors  were fabricated by Bruker BioSpin, S.\ A.\ S.\ of Karlsruhe, Germany, whereas the other custom hardware was constructed at TRIUMF, including the high voltage power supplies for the deflectors. Each of the magnetic elements was mapped by the manufacturer with a Hall effect magnetometer prior to shipment. All of the fringing fields were mapped along with the central fields. The as-built properties of the electromagnetic elements are given in Table \ref{dimensions}.

\subsection{Quadrupole Lenses} \label{quads}

EMMA's first quadrupole lens (Q1) was designed to have high pole tip fields and a short effective length to minimize chromatic aberrations. Its cantilevered support was built so as not to interfere with the placement of 12 of the 16 TIGRESS high purity germanium (HPGe) detectors around the target position while simultaneously keeping the target to Q1 effective field boundary separation small, thereby maximizing angular acceptance. The quadrupoles are arranged in two doublets, Q1-Q2 and Q3-Q4. Both Q2 and Q3 were fabricated according to the same design, while Q1 is smaller; Q4 is larger in order to transmit $m/q$-dispersed ions. Field clamps installed on the upstream and downstream sides of each doublet limit the extent of the fringing fields, which is particularly important given the presence of HPGe detector photomultiplier tubes nearby at the target and focal plane positions. At the factory, the higher multipole components, effective field boundaries, and deviations between the mechanical and magnetic axes were measured with the field clamps in place and found to meet our specifications. 

A transverse Hall probe inserted just below the bore of each quadrupole is used as a reference to set and monitor its field. The Hall effect magnetometers used in all the magnets are model FM-3000-BB-10 Teslameters produced by Projekt Elektronik Mess- und Regelungstechnik GMBH. The field gradient was measured as a function of the reference Hall probe voltage. For each quadrupole, a cylindrical holder with multiple Hall probe positions was precisely machined and aligned, via a laser tracker, to make these measurements. Corrections due to the $<0.1$ mm offset between the mechanical and magnetic axes obtained from the factory field map were neglected. This method produced a calibration that allows the field gradient to be inferred from the reference probe voltage to a precision of $0.1\%$. Accurate calibration of the Hall effect magnetometers was confirmed by comparing with an NMR magnetometer using a uniform dipole reference field.

\subsection{Dipole Magnet} \label{md}

The dipole magnet is a $40^{\circ}$ homogeneous field magnetic sector that was specified to have effective field boundaries (EFBs) inclined by $8.3^{\circ}$ with respect to normal incidence. In order to reduce the horizontal focussing of the sector and increase its vertical focussing, these inclinations are such that ions following trajectories with large bending radii pass through less field and those on small-radius trajectories pass through more field than they would in the case of normal incidence. The poles of the magnet were designed in a three-piece arrangement with small pole edge inserts on either side of a large central piece. These pole edge inserts were machined and re-machined until the measured inclinations of the entrance and exit EFBs were $7.93^{\circ}$ and $8.67^{\circ}$, respectively. The average of the two is $8.3^{\circ}$; the effect of these distinct entrance and exit angles on the overall ion optical performance of the spectrometer was studied using GIOS and found to be compatible with the design requirements after a slight ($<1$~mm) adjustment to the separations between the effective field boundaries of the magnet and the EDs. Each effective field boundary is curved with a radius of 3.47 m to provide a second order correction designed to minimize the $(x|\delta^2_E)$ term. An 8 mm vertical gap between the bottom pole and the vacuum chamber was preserved to allow the placement of thin correction coils on the pole face should they be deemed necessary. A decision to fabricate and install such coils would be made only after a detailed study of ion optical aberrations, which is not currently planned. The field is measured by a Hall probe fixed to the bottom pole piece at a location well within the uniform field region.  

The dipole magnet vacuum chamber has rectangular entrance and exit apertures that measure 205 mm horizontally and 92 mm vertically. In the central region the chamber expands to a maximum horizontal extent of 491 mm. Two straight-through ports aligned with the incoming and outgoing beam axes proved useful when aligning the slit systems located at the entrance and exit of the chamber. Sheets of aluminum honeycomb cores made by Plascore line the vertical walls and bottom surface of the vacuum chamber to minimize scattering of ions on grazing trajectories. The honeycomb cells are 6.35 mm in diameter and the sheets are 3.2 mm thick. 

\subsection{Electrostatic Deflectors}

\begin{figure}[h]
\centering\includegraphics[width=0.75\linewidth, angle=270]{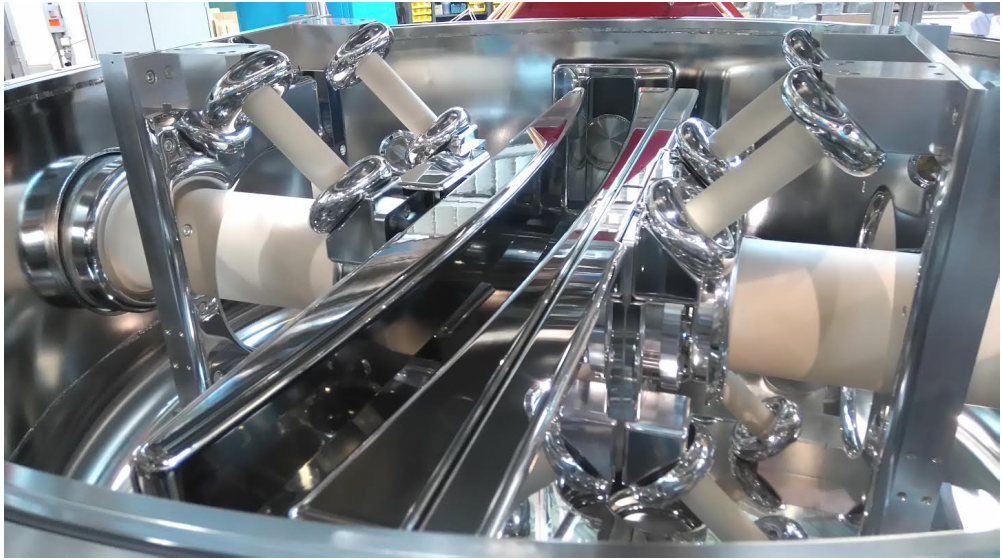}
\caption{\label{edpic}Photograph of the interior of the 2nd electrostatic deflector vacuum vessel.}
\end{figure}

Each of the two electrostatic deflectors (EDs) includes a pair of polished solid titanium electrodes backed by an array of Ti ribs intended to provide structural stability. The ribs and the 25 mm thick electrode of each cathode were machined from a single piece of Ti, as were the 20 mm thick, rib-backed anodes, which were bolstered from behind with an additional ribbed piece of Ti. Each of the four electrodes is supported by four cylindrical ceramic insulators. The insulators were brazed into Ti feet that attach to adjustable, polished Al mounting blocks connected to an Al support frame that rests in a stainless steel vacuum vessel. The vacuum-insulator-conductor triple points are shielded by Al corona rings and all sharp corners and edges are concealed behind polished electrostatic shields fashioned from Al. Figure \ref{edpic} shows the interior of the ED2 vacuum tank. All the components subject to high electric fields were polished extensively using a combination of fine grit sandpaper and a succession of five varying grades of diamond pastes from 9 $\mu$m down to 0.25 $\mu$m grit sizes. For the rounded electrode surfaces that sustain the highest fields, an additional phase of polishing with a colloidal silica suspension was employed.

The electrode support structures were assembled in a clean room to minimize dust deposition on the surfaces. Alignment of the electrode pairs was done to a precision of 0.1 mm inside the clean room using a FARO coordinate measuring machine with an articulating arm and a Leica laser tracker. During the alignment procedure a significant discrepancy was discovered between the measured and specified radii of curvature of the anodes; they are listed for comparison in Table \ref{eds}. The electrodes were positioned such that the gap between them is 125 mm in the centre, which results in a reduction of the gap at the electrode edges compared with the design due to the smaller radii of curvature of the anodes. This manufacturing defect means that the electric fields in the gap are not as homogeneous as specified and that unanticipated second- and higher-order ion optical aberrations are present in the system; it also implies that the first order energy dispersion $(x|\delta_E)$ is not quite as small as planned. These flaws were highly impractical to correct, as the polished, cleaned, and aligned electrodes would have had to be removed, re-machined, and re-polished.

 \begin{table}
 \centering
 \caption{\label{eds}EMMA Electrode Radii of Curvature in mm}
 \begin{tabular}{llll}
 \\
Electrode & ED1 & ED2 & Specification\\
\hline
Anode & 5007.3(55) & 4977.1(55) & 5062.5\\
Cathode & 4953.3(55) & 4952.0(55) & 4937.5\\
\hline
\end{tabular}
\end{table}

High voltage is provided to the electrodes via internal, 40 stage full-wave Cockcroft-Walton DC multipliers built at TRIUMF. The multipliers are driven by external 1~kW Glassman High Voltage Inc.\ PG010K100JD2DRV DC supplies that allow for constant voltage or constant current operation. Each multiplier has been tested without a load to a maximum potential difference of at least 325~kV. To prevent breakdown the multipliers are housed within re-entrant ceramic vessels pressurized with 3 bar of electrically insulating SF$_{6}$ gas. As bias is applied during conditioning, steady state load currents can reach as much as 35~$\mu$A, leading to the emission of X-rays. For shielding purposes, the entire surface of the stainless steel dipole vacuum vessels are covered with 6.35~mm thick lead sheets cut and pounded into various shapes that were affixed using epoxy and contact cement. All the ports and windows are covered with caps containing the same thickness of Pb. Thus far, the first and second electrostatic deflectors have been conditioned to maximum stable potential differences of 340~kV and 440~kV, respectively. We plan to condition them further to a potential difference of 500~kV.

Each electrostatic deflector vacuum tank is pumped by an Agilent TV1001 1000 L/s turbomolecular pump, an Oxford Instruments CryoPlex 8 1500 L/s cryopump, and a Gamma Vacuum TiTan 600L 500 L/s ion pump. The ion pumps run on a diesel-generator-backed uninterruptible electric power circuit in order to ensure high vacuum continuously through electric power bumps and failures. Pressures in the mid $10^{-9}$ Torr range are typically observed with only two pumps running in each isolated vessel.

\section{Experimental Apparatus}

Two 10" OD ConFlat vacuum crosses mounted symmetrically between the electrostatic deflectors and the dipole magnet house horizontal slit systems that can block ions on unwanted trajectories. The two 1.6~mm thick stainless steel plates that make up each slit system can be driven independently to create an opening from 0-180~mm wide. The opening need not be centred on the optic axis. One cross has a 1500 L/s cryopump and the other has a 1000 L/s turbomolecular pump. When either of these pumps is used to evacuate the isolated MD vacuum chamber section a pressure in the low $10^{-9}$ Torr range is typically observed.

Several experimental systems have been designed and installed at EMMA while others are still under construction. These include target and focal plane detector systems and their associated chambers and accessories. Notably, some are being developed with off-site collaborators, such as those at the University of York in the UK.

A 20~cm diameter spherical target chamber couples to the vacuum chamber of the first quadrupole magnet. The chamber is designed to accommodate 12 TIGRESS detectors in a closely packed configuration. It contains a rotary target mechanism that allows for the manual positioning of up to three target foil positions into the beam path. The target chamber also houses an integral, suppressed Faraday cup on a separate rotary actuator; a Ta aperture plate at its entrance lies in the same plane as the target foils. The Ta plate has a 1~mm diameter aperture through which the beam is tuned by maximizing the current on the Faraday cup while minimizing that on the aperture plate.

The target chamber has provisions for mounting two 150~mm$^2$ silicon surface barrier detectors centred at 20$^{\circ}$ angles with respect to the beam axis downstream of the target. They are used to monitor the beam flux and target condition via elastic scattering and are primarily intended for the normalization of reaction cross section measurements. Additionally, a highly-segmented, annular silicon detector can be mounted 33~mm upstream or downstream of the target position to detect light charged particles. A thin C foil can be positioned 116~mm downstream of the target position in order to restore the charge state distributions of transmitted ions to equilibrium through charge-changing collisions following possible internal conversion decays. Similarly, an energy degrader foil can be mounted 68 mm downstream of the target. To exclude ions on trajectories outside of the angular acceptance of the spectrometer, a circular aperture at the exit of the target chamber defines a cone of half-angle $4.2^{\circ}$ and a solid angle of 17~msr; an additional, optional $\pm2^{\circ}$ aperture can be used to restrict the horizontal angular acceptance when high $m/q$ resolving power is required, as described in Ref.\  \cite{davids05}.

A separate target chamber designed to accommodate highly-segmented rectangular and annular silicon detectors as well as 12 TIGRESS detectors, dubbed SHARC-II, has been designed at the University of York and is in the final stages of fabrication. It can be positioned so that the target is separated from the Q1 effective field boundary by the standard 25 cm.

The focal plane station has a modular design in which detectors may be inserted and removed easily according to the experimental requirements. A position-sensitive parallel grid avalanche counter (PGAC) and an energy-sensitive ionization chamber (IC) make up the standard complement of focal plane detectors. They are mounted in separate vacuum chambers that can be joined together, allowing for the use of the PGAC without the IC if only $m/q$ information is required. In the current configuration, the installation of the PGAC is mandatory while the IC is optional. A 3000~mm$^2$ ion-implanted Si detector can be mounted directly behind the PGAC, as can a double-sided Si strip detector. The latter can also be mounted behind or inside the IC. All of the detectors are read out using a version of the MIDAS data acquisition system.

Both the PGAC and the IC are filled with isobutane as an ionization medium and have an active area of 154 mm by 54 mm, with a larger extent in the horizontal, dispersive direction than in the vertical direction. The PGAC, which measures 73 mm between entrance and exit foils, operates at pressures between 2 and 4 Torr while the 40~cm long, 16~anode segment IC is designed to operate at pressures of $10-100$ Torr. Provisions have been made to mount up to 4 clover-shaped HPGe detectors of the type used in the GRIFFIN spectrometer \cite{garnsworthy18} at the focal plane to study isomers and delayed activities.

The focal plane station has a set of 4 independently actuated 1.6~mm thick stainless steel plates just upstream of the PGAC that together constitute a slit system. Two of the plates are mounted on combined rotary and linear motion feedthroughs that present different profiles to the incident ions depending on their angle. In this way the focal plane slit system can obscure the entire focal plane except for $1-3$ openings of continuously variable position and width, enabling the simultaneous transmission of up to 3 charge states.

\section{Initial Measurements}\label{measurements}

\begin{figure}[h]
\centering\includegraphics[width=\linewidth]{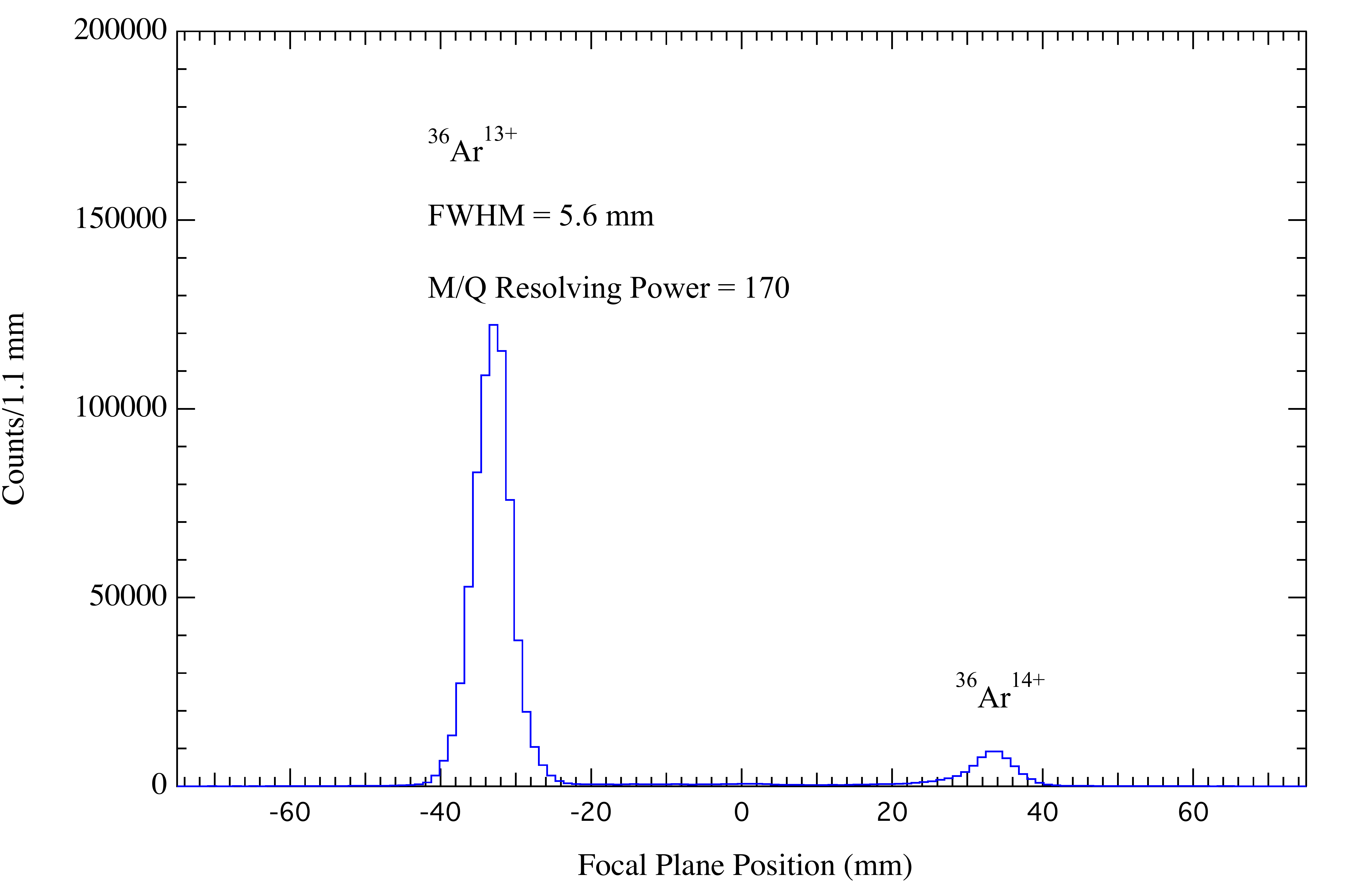}
\caption{\label{firstspectrum}First $m/q$ spectrum obtained with EMMA; it resulted from bombarding a 4.5 $\mu$m Au foil with an 80~MeV $^{36}$Ar beam. Multiply-scattered Ar ions filled both the angular and energy acceptances of the spectrometer. The two detected charge states are separated by the 66~mm expected from the tuned $m/q$ dispersion of 10~mm/\%.}
\end{figure}

Due to a flurry of last minute activity on the focal plane station vacuum and control systems just prior to its first scheduled beam time, there was no opportunity to first test the spectrometer with an alpha source. Hence initial focussing and dispersion tests were carried out by bombarding a 4.5~$\mu$m thick Au foil with an 80~MeV $^{36}$Ar beam. The spectrometer was initially set to transmit multiply scattered, 19~MeV $^{36}$Ar$^{13+}$ ions. The foil was sufficiently thick that the multiply scattered Ar ions filled the angular and energy acceptances of the spectrometer. After observing a single, well-defined $m/q$ peak we set the fields for the same energy but charge state 13.5$^+$ and transmitted the 13$^+$ and 14$^+$ charge states simultaneously. As shown in Figure \ref{firstspectrum}, the peaks of the two charge states were separated by the expected 66~mm distance corresponding to the design $m/q$ dispersion of 10 mm/\%. The different peak heights reflect the charge state distribution of the transmitted $^{36}$Ar ions.

Following this initial in-beam test, we carried out a series of ion optical studies with a $^{148}$Gd $\alpha$ source to study the angular focussing, $m/q$ dispersion, and energy dispersion cancellation. These tests will be described in a forthcoming publication. The spectrometer was further tested when we bombarded a 900~$\mu$g~cm$^{-2}$ natural Cu target foil with both stable and radioactive Na beams and set the spectrometer to detect fusion products. In these measurements the $\pm2^{\circ}$ horizontal aperture was in place, the PGAC was operated at a pressure of 2 Torr, and the 3000~mm$^2$ Si detector was positioned directly behind the PGAC to aid in distinguishing fusion products from scattered beam ions. We first bombarded the Cu target with an 84~MeV $^{23}$Na beam from the ISAC offline ion source \cite{jayamanna14} and set the spectrometer for 18.4~MeV, 81~u, $10^+$ recoils.

\begin{figure}[h]
\centering\includegraphics[width=\linewidth]{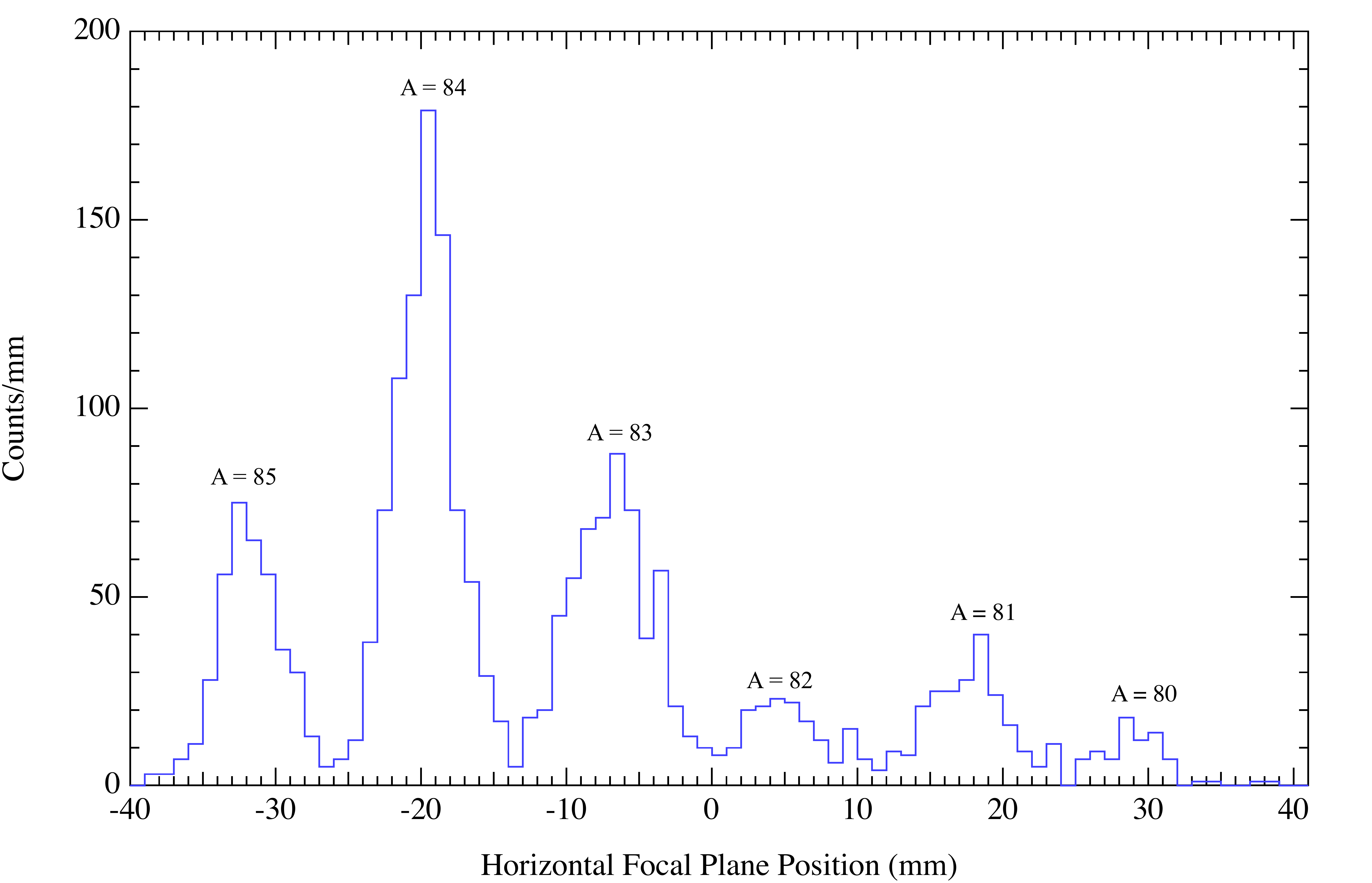}
\caption{\label{ribmassspectrum}First $m/q$ spectrum obtained using EMMA to detect products of reactions induced by a radioactive beam. A 900~$\mu$g~cm$^{-2}$ natural Cu foil was bombarded with an 87~MeV $^{24}$Na beam and the spectrometer was tuned to transmit 17.1~MeV, 82~u, $10^+$ recoils. This position spectrum from the PGAC was gated to include only events with an energy signal in the Si detector corresponding to fusion residues.}
\end{figure}

After we obtained $m/q$ spectra of the fusion products with several overlapping spectrometer settings, the operations group delivered an 87~MeV, radioactive $^{24}$Na beam produced in a SiC target by TRIUMF's 500~MeV proton beam and doubly ionized by a forced electron beam induced arc discharge ion source \cite{bricault14}. The operators succeeded in tuning more than 90\% of the radioactive beam through the 1~mm aperture of EMMA's Faraday cup. Beam intensities on target ranged from 1 to $4\times10^7$~s$^{-1}$ and spectra were measured at several field settings. Figure \ref{ribmassspectrum} shows that obtained when the spectrometer was set for 17.1~MeV, 82~u, $10^+$ recoils. Beam suppression at these settings exceeded a factor of $10^9$ and the $m/q$ resolving power was measured to be as large as 240 (FWHM), which is more than adequate to resolve masses around $A=90$.

\section{Status and Future Plans}\label{plans}

High voltage conditioning is expected to be completed within the next 2 months, at which point the spectrometer will have been fully commissioned. Four EMMA experiments and two letters of intent have been approved by the TRIUMF subatomic physics experiment evaluation committee. All are motivated by nuclear astrophysics and all but one of them require the installation of the TIGRESS $\gamma$-ray spectrometer around the EMMA target position to provide spectroscopic information. This installation is planned to be completed in the spring of 2019, at which point the EMMA scientific program can begin in earnest.

\section{Acknowledgements}

BD acknowledges generous support from the Natural Sciences and Engineering Research Council of Canada and is grateful to F.~Cifarelli for his mechanical design work. OSK acknowledges support from the Villum Foundation. TRIUMF receives federal funding via a contribution agreement through the National Research Council of Canada. The UK authors acknowledge the support of the Science and Technology Facilities Council.




\bibliographystyle{model1-num-names}
\bibliography{emma}







\end{document}